\documentstyle[11pt,paspconf,epsf]{article}

\begin{document}

\title{B0218+357: Time Delays and New MERLIN/VLA\\ 
5~GHz Maps of the Einstein Ring}
\author{A.D. Biggs, I.W.A. Browne, P.N. Wilkinson, and T.W.B. Muxlow}
\affil{University of Manchester, Nuffield Radio Astronomy Laboratories,
Jodrell Bank, Macclesfield, Cheshire SK11 9DL, UK}
\author{P. Helbig and L.V.E. Koopmans}
\affil{University of Groningen, Kapteyn Astronomical Institute, Postbus 800,
9700 AV Groningen, the Netherlands}

\begin{abstract}
This poster presents a new 5~GHz combined MERLIN/VLA map of B0218+357 which
shows for the first time believable substructure in the Einstein ring. This
will now be exploited for further constraints on the model which presently 
dominates the error on the estimate of $H_0$ derived from the time delay
(10.5$\pm$0.4 days) measured for this system.
\end{abstract}


\section{Introduction}

Upon its discovery in 1992, the lens system B0218+357 (Fig. 1, left) was 
immediately recognised as an excellent candidate for determining $H_0$. The 
source is radio loud, highly polarised and variable and both redshifts are 
known. Modelling the mass responsible for the lensing is relatively simple 
compared to many other lenses as the deflector is believed to be an isolated 
face-on spiral galaxy. The Einstein ring is potentially a particularly useful 
source of modelling constraints as it effectively samples the lensing 
potential over many lines-of-sight (Kochanek 1990).

\section{Time Delay Summary}

The time delay has been measured to be 10.5$\pm$0.4 days at 95\% confidence
(Biggs et al. 1999) from VLA monitoring of total flux density and polarisation 
at two frequencies, 8.4 and 15 GHz. Although the time delay for this system is 
in little doubt, the above value has been independently confirmed (Cohen et al.
1999).

Modelling of this system is still at an early stage, but an initial 
Singular Isothermal Ellipsoid mass model constrained using the observed VLBI 
substructure of the two compact images and the flux density ratio found from 
the VLA monitoring, when combined with the above value of the time delay,
gives an $H_0$ of 69$^{+13}_{-19}$\,km\,s$^{-1}\,$Mpc$^{-1}$ (95\%
confidence).

\section{MERLIN/VLA 5~GHz Map}

To date, no models have exploited the Einstein ring for constraints on 
the lensing due to a lack of resolution and sensitivity in existing radio 
maps. A new image made from combined 5~GHz VLA/multi-frequency synthesis 
MERLIN data is shown in Fig. 1 (right). The MERLIN data give high resolution 
($\sim$50 mas) whilst the many short baselines of the VLA provide much more 
sensitivity and aperture coverage for detecting and mapping the extended ring 
emission. 

\begin{figure}
\plotfiddle{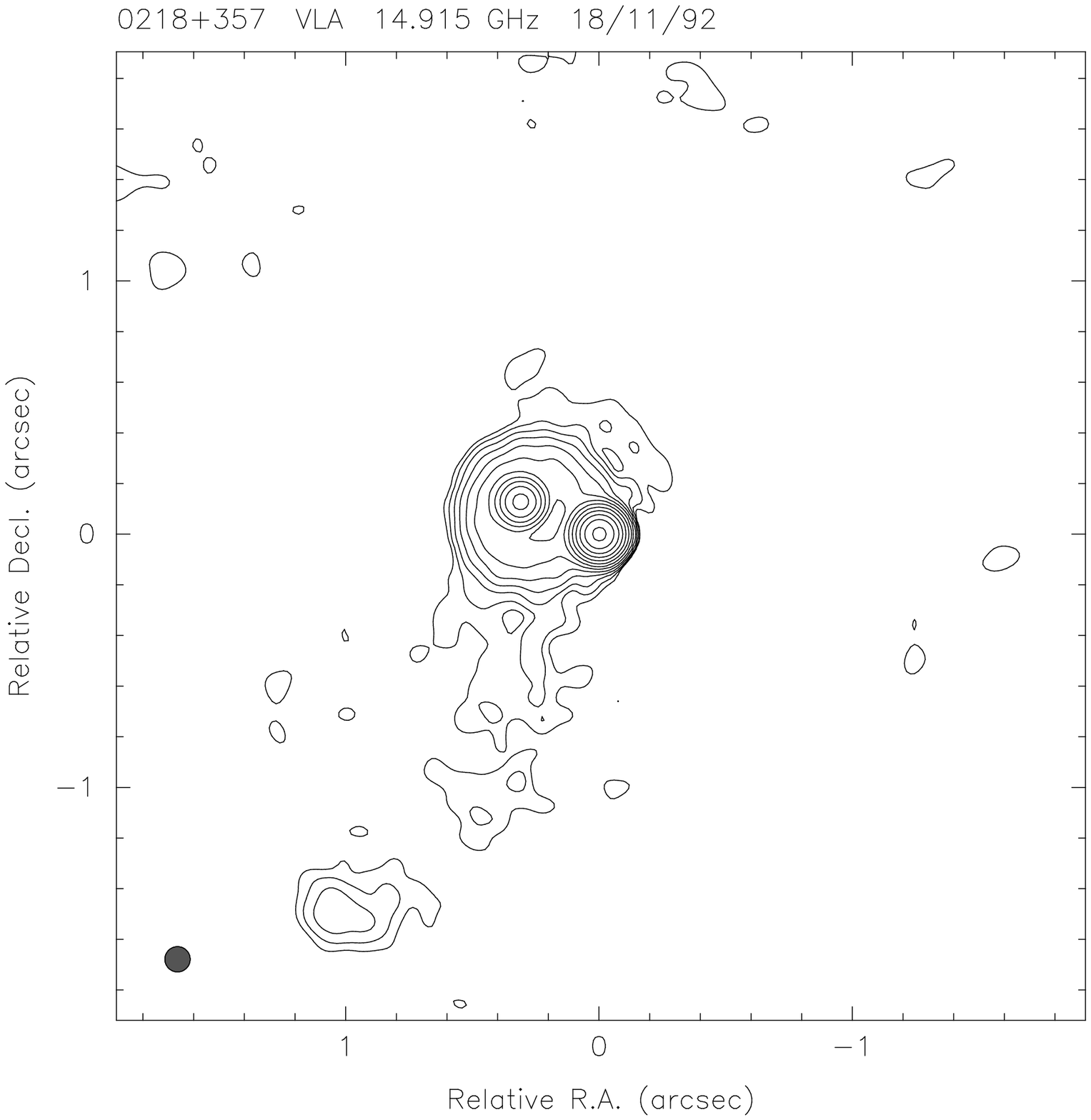}{4cm}{0}{30}{30}{-200}{-75}
\plotfiddle{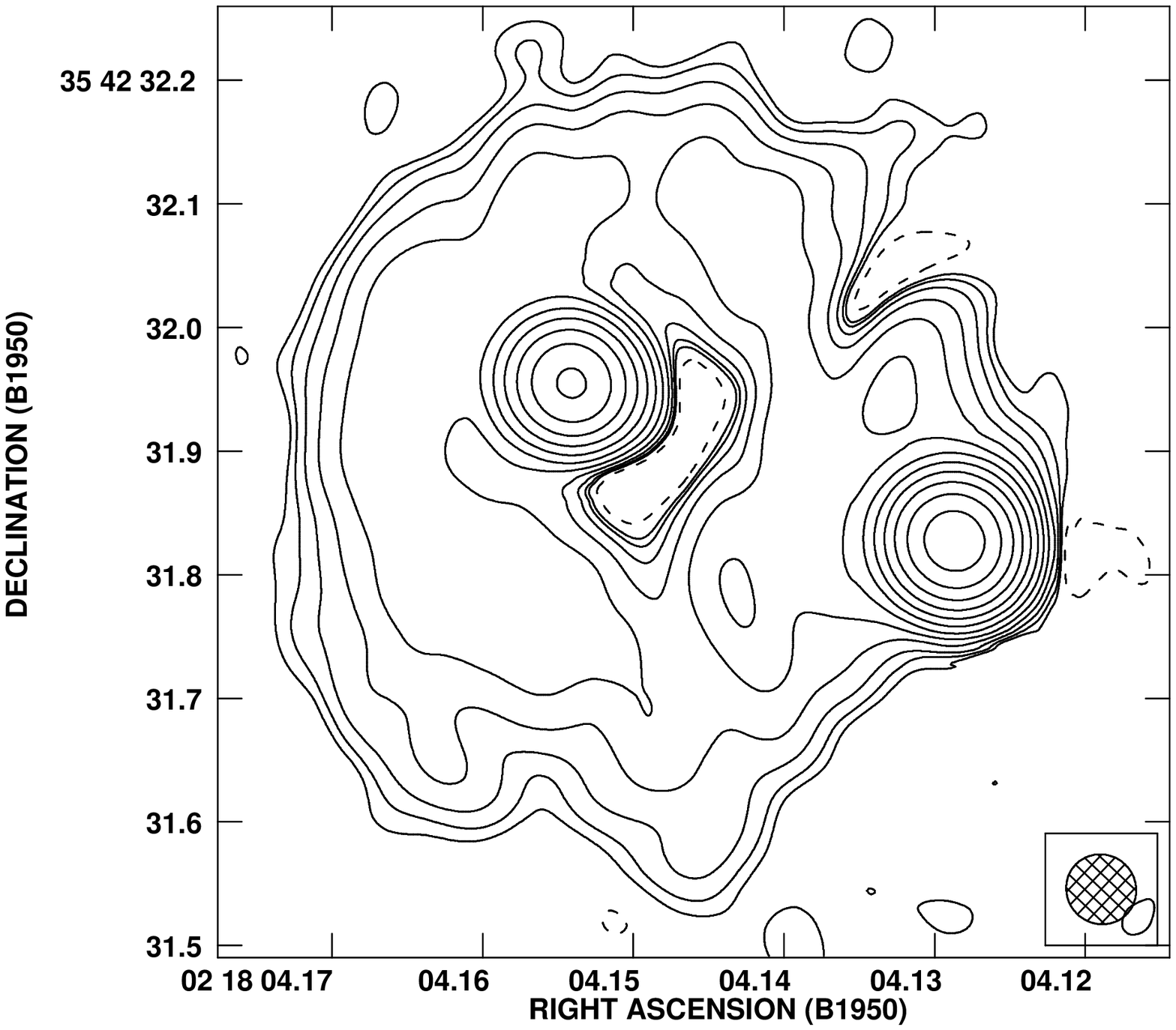}{0cm}{0}{34}{34}{-20}{-58}
\caption{VLA 15~GHz map (left), MERLIN/VLA 5~GHz map (right).}
\end{figure}

The most prominent feature revealed by the new map is the hole in the 
centre of the ring, a hint of which is also seen in the 15~GHz VLA image. The 
MERLIN 5 GHz map also shows valleys of reduced surface brightness stretching 
away north and south from the hole that make the ring's morphology more akin 
to that of two arcs. Each of these is further separated into several discrete 
areas of increased brightness. Work is at present underway to exploit this 
new image for extra modelling constraints.

\acknowledgments

MERLIN is a national UK facility operated by the University of Manchester on 
behalf of PPARC. The VLA is operated by the National Radio Astronomy 
Observatory, which is a facility of the National Science Foundation operated 
under cooperative agreement by Associated Universities, Inc. ADB acknowledges 
the receipt of a PPARC studentship. This research was supported in part by 
European Commission, TMR Programme, Research Network Contract ERBFMRXCT96-0034 
``CERES''.

\end{document}